\documentclass[12pt]{JHEP3}

\usepackage{mathrsfs}
\usepackage{amsmath,amssymb}
\usepackage{epsfig}
\usepackage{relsize}
\usepackage{graphicx}
\input epsf

\usepackage{relsize}
\usepackage{epsfig}

\input epsf

\usepackage{mathrsfs}
\usepackage{amsmath,amssymb}
\usepackage{epsfig}
\input epsf

\def\x'{\mathaccent 19 x}
\def\y'{\mathaccent 19 y}
\def\n'{\mathaccent 19 n}
\def\u'{\mathaccent 19 u}

\def\et'{\mathaccent 19 \eta}
\def\th'{\mathaccent 19 \theta}
\def\lam'{\mathaccent 19 \lambda}
\def\varet'{\mathaccent 19 \vartheta}
\def\rh'{\mathaccent 19 \rho}
\def\ph'{\mathaccent 19 \phi}
\def\xb'{\mathaccent 19 {\bar{x}}}

\def \A {{\cal{A}}}

\def\det{\hbox{det}}
\def\be{\begin{equation}}
\def\ee{\end{equation}}

\newcommand{\bea}{\begin{eqnarray}}
\newcommand{\eea}{\end{eqnarray}}

\def\a {\alpha}
\def\b {\beta}

\def\pa {\partial}

\def\k{\kappa}

\newcommand{\alg}[1]{\mathfrak{#1}}

\newcommand{\AdS}{{\rm  AdS}_5\times {\rm S}^5}

\newcommand{\sfrac}[2]{{\textstyle\frac{#1}{#2}}}

\def\CP{\mathbb{ CP}}

\newcommand{\ads}{{\rm  AdS}}

\def \k {\kappa}

\author{Marcin Dukalski $^{a}$
\  and Stijn J. van Tongeren $^{a}$ \footnote{email: M.S.Dukalski@students.uu.nl, S.J.vanTongeren@students.uu.nl} {}
 \\ $^{a}$ {\it Institute for Theoretical
Physics and Spinoza Institute,\\ ~~Utrecht University, 3508 TD
Utrecht, The Netherlands}}

\title{Fermionic reductions of the ${\rm AdS}_4\times \mathbb{CP}^3$ superstring}

\preprint{
          \tiny{ITP-UU-09-21}\\[-.5ex]
          \tiny{SPIN-09-20}\\[-.5ex]
          }

\maketitle

\abstract{We discuss fermionic reductions of type IIA superstrings
on  ${\rm AdS}_4\times \mathbb{CP}^3$ in relation to the
conjectured ${\rm AdS}_4/{\rm CFT}_3$ duality. The  superstring
theory is described by means of a coset model construction, which
is classically integrable. We discuss the global light-cone
symmetries of the action and related $\k$-symmetry gauge choices,
and also present the complete quartic action in  covariant form
with respect to these. Further, we study integrable (fermionic)
reductions, in particular, a reduction yielding a quadratic action
of two complex fermions on the string world-sheet. Interestingly,
this model appears to be exactly the same as the corresponding
integrable reduction found in the $\AdS$ case.}

\newpage

\begin{document}



\section{Introduction}
Recently Aharony, Bergman, Jafferis and Maldacena conjectured that
${\cal N}=6$ supersymmetric Chern-Simons theory in three
dimensions has a holographic dual which at strong coupling can be
effectively described by type IIA superstings  on ${\rm
AdS}_4\times \mathbb{CP}^3$ background \cite{Aharony:2008ug}.

\smallskip

It appears that the Green-Schwarz superstring on the ${\rm
AdS}_4\times \mathbb{CP}^3$ background admits a description in
terms of a coset sigma model, quite analogous to strings on
$\AdS$. Normally the Green-Schwarz action contains 32 fermions,
only half of which are physical due to $\k$-symmetry, whereas this
coset model, by construction, contains 24 fermions. This apparent
mismatch can be understood by considering the coset model to be a
partially $\k$-symmetry fixed version of the Green-Schwarz action,
and, in accordance, the coset model possesses local fermionic
symmetry that reduces the number of physical fermions to 16. The
${\rm AdS}_4\times \mathbb{CP}^3$ background is diffeomorphic to
the coset space ${\rm SO}(3,2)/{\rm SO}(3,1)  \times {\rm
SO}(6)/{\rm U}(3)$. The coset numerator ${\rm SO}(3,2)   \times
{\rm SO}(6)$ is the bosonic subgroup of ${\rm OSP}(2,2|6)$, which
suggests that fermions can be included in the theory by
considering the coset space  ${\rm OSP}(2,2|6)/\left({\rm SO}(3,1)
\times {\rm U}(3)\right)$.

\smallskip

An important advantage of the coset sigma model construction is
that it exhibits manifest integrability. Analogously to the $\AdS$
case \cite{arutyunov-2009}, classical integrability of this model
can be concluded immediately from the existence of the
corresponding Lax pair \cite{arutyunov-2008,Stefanski:2008ik}; the
corresponding algebraic curve encoding solutions of classical
equations of motion has been obtained in \cite{Gromov:2008bz}. We
also note that recently the full Green-Schwarz action for strings
on ${\rm AdS}_4\times \mathbb{CP}^3$ has been constructed, but the
integrable properties of this formulation still remain an open issue \cite{Gomis:2008jt,Grassi:2009yj}. The pure spinor formulation for this background has been explored in \cite{Bonelli:2008us}.

\smallskip

The question of quantum integrability is rather subtle,
especially, taking into account that classical integrability of
the $\mathbb{CP}^3$ bosonic model is known to be spoiled by
quantum corrections. Following an analogous construction for  the
$\AdS$, and the leading perturbative result found in field theory
\cite{MZ} (see also \cite{Bak:2008cp}-\cite{Bak:2009mq}), an
all-loop asymptotic Bethe ansatz has been proposed in
\cite{Gromov:2008qe}. It would be important to find further
support for its validity from direct field-theoretic computations
on the string world-sheet, following the lines, e.g., in
\cite{McLoughlin:2008ms}-\cite{Astolfi:2008ji}.

\smallskip

In order to shed some more light on the integrability of
superstrings on  the ${\rm AdS}_4 \times \mathbb{CP}^3$
background, in this note we will investigate possible consistent
truncations of the string sigma model. This problem has already
received some attention in the bosonic sector of the theory
\cite{Rashkov:2008rm}, but since fermions are expected to play an
important role in maintaining quantum integrability, we will
consider consistent {\it fermionic} reductions of the superstring
Lagrangian.

\smallskip

Our treatment starts with describing a particular parametrization
of the coset element ${\rm SO}(3,2)/{\rm SO}(3,1) \times {\rm
SO}(6)/{\rm U}(3)$ suitable for the light-cone gauge fixing
\cite{AF}. We then identify the  manifest bosonic symmetry group
of the light-cone Lagrangian, which appears to be in agreement
with earlier findings in
\cite{sundin-2008,Bykov:2009jy,Zarembo:2009au}. Further we
determine the specific form of the $\k$-symmetry parameter for our
current coset parametrization based on the analysis of
$\k$-symmetry presented in \cite{arutyunov-2008}. We then proceed
to fix a $\k$-symmetry gauge compatible with these bosonic
symmetries. Next we present the complete sigma model Lagrangian in
the expansion up to quartic order in fields,
 given in a manifestly
covariant form with respect to the manifest bosonic symmetries.
This Lagrangian can be used to compute the full tree level
world-sheet S-matrix. Finally, we discuss a consistent truncation
of the full Lagrangian down to a model containing two complex
fermions which, being a consistent truncation, maintains
integrability inherited from the full model. Interestingly, the
emerging reduced model appears to be completely equivalent to an
integrable system arising under the corresponding consistent
truncation of the $\AdS$ superstring
\cite{Alday:2005jm,Arutyunov:2005hd}.

\smallskip

\section{Sigma-model Lagrangian}
The coset sigma model is based on a coset element that parametrizes a space, such that the fields take values in a non-linear way on a chosen manifold, and does not require the knowledge of the Green-Schwarz action.
In particular the ${\rm AdS}_4 \times \mathbb{CP}^3$ coset sigma model is constructed on the coset space
\begin{eqnarray*}\label{coset}
\frac{{\rm OSP}(2,2|6)}{{\rm SO}(3,1) \times {\rm U}(3)},
\end{eqnarray*}
where we note that   ${\rm USP}(2,2)$, a bosonic subgroup of $\rm OSP(2,2|6)$, is locally isomorphic to  $\rm SO(3,2)$. The complete construction of this model is discussed in \cite{arutyunov-2008}.
\subsection{Superalgebra $\mathfrak{osp}(2,2|6)$ and ${\mathbb Z}_4$-grading}
An arbitrary element belonging to the $\mathfrak{osp}(2,2|6)$ superalgebra is given by a $10 \times 10$ matrix
\begin{equation*}
A=\left(\begin{array}{ll} X ~&~ \theta \\ \eta ~&~ Y   \end{array}\right),
\end{equation*}
where $X$ and $Y$ are $4 \times 4$ and $6 \times 6$ matrices containing bosonic fields, and $\eta$ and $\theta$ are $6 \times 4$ and $4 \times 6$ blocks
containing fermionic  fields respectively, where $A$ has to obey the following two conditions
\begin{eqnarray}
&A^{st}&\left(\begin{array}{cc} C_{4} ~~~&~~~ 0 \\ 0 ~~~&~~~
{\mathbb I}_{6\times 6}
\end{array}\right) + \left(\begin{array}{cc} C_{4} ~~~&~~~ 0 \\ 0
~&~ {\mathbb I}_{6\times 6} \end{array}\right)A=0 \label{transcon}\, ,\\
 &A^{\dagger}&\left(\begin{array}{cc} \Gamma^0 ~&~ 0 \\ 0 ~&~
-{\mathbb I}_{6\times 6}
\end{array}\right) + \left(\begin{array}{cc} \Gamma^0 ~&~ 0 \\ 0
~&~ -{\mathbb I}_{6\times 6} \end{array}\right)A=0 \label{realcon}\, .
\end{eqnarray}
Here $C_4$ denotes an arbitrary real, skew symmetric, charge conjugation matrix satisfying $C_4^2=-\mathbb{I}$. For purpose of this paper we have chosen $C_4=i \Gamma_0 \Gamma_3$. Additionally
we define the supertranspose of a matrix to be
\begin{eqnarray}
\nonumber A^{st}=\left(\begin{array}{rr} X^t ~&~-\eta^t \\
 \theta^t ~&~ Y^t   \end{array}\right) \, .
\end{eqnarray}
For the odd blocks of $A$, equations (\ref{transcon}) and (\ref{realcon}) imply the following fermionic transposition and reality conditions,
\begin{eqnarray}
\nonumber \eta=-\theta^t C_4\, ,~~~~~~~~\theta^*=i\Gamma^3 \theta\, .
\end{eqnarray}
These reduce the number of fermionic degrees of freedom from 48 complex, down to 24 real.
This algebra possesses a fourth order outer automorphism
\begin{eqnarray}
\nonumber  \Omega(A)&=& \Upsilon A \Upsilon^{-1}\, ,
\end{eqnarray}
where the matrix $\Upsilon$ is given in Appendix \ref{GandTmatrices}.
This gives the algebra a $\mathbb{Z}_4$ grading
\begin{eqnarray}
\nonumber \mathcal{A}=\mathcal{A}^{(0)}\oplus \mathcal{A}^{(1)}\oplus \mathcal{A}^{(2)}\oplus \mathcal{A}^{(3)}\, ,
\end{eqnarray}
such that $[\mathcal{A}^{(k)},\mathcal{A}^{(m)}]\subseteq\mathcal{A}^{(k+m)}$ modulo $\mathbb{Z}_4$, and where the subspace $\mathcal{A}^{(k)}$
 is an eigenspace of the map $\Omega$
\begin{eqnarray}
\nonumber \Omega(\mathcal{A}^{(k)})=i^{k}\mathcal{A}^{(k)} \, .
\end{eqnarray}
In particular, the stationary subalgebra of
$\Omega$ is determined by the conditions
\begin{eqnarray}
\nonumber [\Gamma^5,X]=0\, ,~~~~~~~~[K_6,Y]=0 ,
\end{eqnarray}
where $\Gamma^5=-i\Gamma^0\Gamma^1\Gamma^2\Gamma^3=K_4C_4$ is a matrix defined in terms of anticommuting $\Gamma^i$ matrices used to parametrize the ${\rm AdS}_4$ space.  The space $A^{(2)}$ is spanned by matrices satisfying
\begin{eqnarray}
\nonumber \Omega(A)=\Upsilon A \Upsilon^{-1}=-A\, .
\end{eqnarray}
This implies the following conditions
\begin{eqnarray}
\nonumber  \{X,\Gamma^5\}=0\,~~~~~~~~\{Y,K_6\}=0 \, ,
\end{eqnarray}
which are solved by
\begin{eqnarray}
\nonumber X=x_{\mu}\Gamma^{\mu},~~~~~~~ Y=y_i T_i \, ,
\end{eqnarray}
providing a parametrization of the coset space of
${\rm AdS}_4=\tfrac{{\rm SO(3,2)}}{{\rm SO}(3,1)}$ and $\mathbb{CP}^3=\tfrac{{\rm SO(6)}}{{\rm U}(3)}$ respectively. The two sets of matrices
$\Gamma^{\mu}$ and $T_i$ are given in Appendix \ref{GandTmatrices}.

\subsection{The Lagrangian}
To further progress in the construction of the Lagrangian, let us use an element of the coset $g$ to
build the following current (one-form)
\begin{eqnarray}
\nonumber A=-g^{-1}{\rm d}g=A^{(0)}+A^{(2)}+A^{(1)}+A^{(3)}\, ,
\end{eqnarray}
where on the right hand side we exhibited its  $\mathbb{Z}_4$-decomposition. By
construction $A$ has vanishing curvature
\begin{eqnarray}
\nonumber \partial_{\alpha} A_{\beta}-\partial_{\beta} A_{\alpha}-[A_{\alpha},A_{\beta}]=0\, .
\end{eqnarray}
The sigma model is given by the following action
\begin{eqnarray}
\nonumber S= -\frac{\sqrt{\lambda}}{4\pi}\int {\rm d} \tau {\rm d} \sigma \mathcal{L},
\end{eqnarray}
where $\lambda$ is the 't Hooft coupling constant, related to the AdS radius as $\sqrt{\lambda}=R^2 / \alpha'$.
The Lagrangian density $\mathcal{L}$ is the sum of the kinetic and the Wess-Zumino terms
\begin{eqnarray}
\label{lagrangianA2A2plusA1A3}
\mathcal{L} =\gamma^{\alpha\beta}{\rm str}\big(A^{(2)}_{\alpha}A^{(2)}_{\beta}\big)+\kappa
\epsilon^{\alpha\beta}{\rm str}\big(A^{(1)}_{\alpha}A^{(3)}_{\beta}\big)\, .
\end{eqnarray}
where ${\rm str}$ denotes the supertrace, $\gamma^{\alpha\beta}$ is the Weyl-invariant
world-sheet metric with $\det\gamma=-1$, and  $\epsilon^{\tau\sigma}=1$.The parameter $\kappa$ in front of the
Wess-Zumino term is kept arbitrary, however the requirement of $\k$-symmetry and   integrability will  fix it to $\kappa=\pm 1$.

 The equations of motion derived from (\ref{lagrangianA2A2plusA1A3}) are
\begin{equation}\label{EOMmatrix}
\pa_\a{\rm \Lambda}^{\a}-[A_{\a},\Lambda^{\a}]=0\, ,
\end{equation}
where we have defined,
\begin{eqnarray}
\nonumber \Lambda^{\a}=\gamma^{\a\b}A^{(2)}_{\b} -\sfrac{1}{2}\kappa\,\epsilon^{\a\beta}(A^{(1)}_{\beta}-A^{(3)}_{\beta}).
\end{eqnarray}
For further details see \cite{arutyunov-2008}.

\subsection{Kappa symmetry}

Kappa symmetry is a well known local fermionic symmetry in string theory. In the case of type IIA Green-Schwarz string theory it typically allows one to gauge away 16 of the 32 fermionic degrees of freedom. As was argued in \cite{arutyunov-2008}, our coset model is to be understood as a Green-Schwarz string theory with $\k$-symmetry \emph{partially} fixed. In accordance with this picture, there is a local fermionic symmetry in this coset model that allows one to gauge away precisely the eight fermionic degrees of freedom needed.

\smallskip

Contrary to the global bosonic symmetries which are realized by multiplication from the left, $\k$-symmetry is here understood as the right local action of a fermionic element $G= \mbox{exp} \hspace{2pt}\epsilon$ from $\mbox{OSP}(2,2|6)$ on our coset representative $g$ \cite{arutyunov-2008}:\begin{equation*} g G(\epsilon) = g^\prime g_c \end{equation*}
where $\epsilon\equiv \epsilon(\tau,\sigma)$ is a local fermionic parameter, and $g_c$ is a compensating element from $\mbox{SO}(3,1)\times \mbox{U}(3)$. The action is not invariant under arbitrary variations of this form. In order for this to be the case, the fermionic parameter has to be of a form presented below, and the whole transformation will have to be accompanied by a transformation of the metric.

\smallskip

The fermionic parameter $\epsilon$ can of course be decomposed into an element of degree one, $\epsilon^{(1)}$, and one of degree three, $\epsilon^{(3)}$. The transformation above is a symmetry of the action, provided the following form of the $\kappa$-symmetry parameter $\epsilon^{(1)}$
\begin{eqnarray}
\hspace{-0.7cm}\epsilon^{(1)}&=&A_{\a,-}^{(2)}A_{\b,-}^{(2)}\kappa_{++}^{\a\b}+\kappa_{++}^{\a\b}A_{\a,-}^{(2)}A_{\b,-}^{(2)}+A_{\a,-}^{(2)}\kappa_{++}^{\a\b}A_{\b,-}^{(2)}\nonumber\\
\label{eq:kappapar}
&&-\frac{1}{8}\,{\rm str}(\Sigma A_{\a,-}^{(2)}A_{\b,-}^{(2)})\kappa_{++}^{\a\b}\, ,\label{kappa}
\end{eqnarray}
where $\kappa_{++}^{\a\b}$ is the $\kappa$-symmetry parameter which is assumed to be independent of the dynamical fields of the model. Obviously, $\kappa^{\a\b}_{++}$ must be an element of $\alg{ osp}(2,2|6)$, and $\epsilon^{(1)}\in \A^{(1)}$ provided $\kappa^{\a\b}_{++}$ is. The form of the $\kappa$-symmetry parameter $\epsilon^{(3)}$ is analogous. The accompanying variation of the metric is then given by \begin{eqnarray*} \delta \gamma^{\a\beta}=\frac{1}{2}\, {\rm str}\Big(\Sigma A^{(2)}_{\delta,-} [\kappa_{++}^{\a\b},A^{(1),\delta}_+]\Big)+\frac{1}{2}\, {\rm str}\Big(\Sigma A^{(2)}_{\delta,+} [\varkappa_{--}^{\a\b},A^{(3),\delta}_-]\Big)\, , \end{eqnarray*} where $\varkappa_{--}^{\a\b}\subset \A^{(3)}$ is another independent $\kappa$-symmetry parameter that comes in with $\epsilon^{(3)}$. Finally,  along with the above transformations, the parameter $\kappa$ in the Lagrangian is required to be equal to plus or  minus one to guarantee invariance of the action. For details on the derivation of $\kappa$-symmetry and the explicit form of the $\kappa$-symmetry parameter $\epsilon$, see \cite{arutyunov-2008}.

\smallskip

This symmetry can be used to gauge away eight fermionic degrees of freedom. The question then is what the best suited gauge choice is, or equivalently, what the most convenient form of the gauge fixed fermionic coset element is. In the next sections we will discuss the manifest bosonic symmetry of the light-cone gauge fixed action for our choice of coset parametrization\footnote{The precise form of the $\kappa$-symmetry parameter is affected by this choice of parametrization, as discussed in the Appendix.}, and its action on the bosonic and fermionic fields. Ideally one would like to choose a $\kappa$-symmetry gauge that is preserved under the action of the complete manifest bosonic symmetry group. We will present one such choice, directly after discussing the bosonic symmetries of course.

\section{$\kappa$-gauge fixed Lagrangian}
\subsection{Coset parametrization, symmetries and $\kappa$-gauge
choice}\label{cosetsymetriesandkappa}

In anticipation of imposing a uniform light-cone gauge, with the
light-cone coordinates taken to be combinations of the $\ads$ time
variable $t$ and the $\CP^3$ spherical coordinate angle $\phi$, we
take the following choice of coset parametrization
\begin{equation}
\label{eq:cospar} g=\Lambda (t,\phi) g_\chi g_B,
\end{equation}
\noindent where
\begin{eqnarray*}
\Lambda (t,\phi) = \left(
\begin{array}{ll}
 e^{\frac{i}{2} t \Gamma^0} & 0 \\
 0 & e^{-\frac{\phi}{2}\left(T_{34}+T_{56}\right)}
\end{array}
\right), ~~\, g_B=\left(\begin{array}{l r} e^{i x_i \Gamma^i} &
0\\ 0& e^{i y_j T_j}
\end{array}\right)\,,
\end{eqnarray*}
and
\begin{eqnarray*}
~~ g(\chi)=\exp(\chi), ~~{\rm for} ~~ \chi=\left(\begin{array}{l
r} 0 & \kappa\\ -\kappa^t C_4& 0
\end{array}\right).
\end{eqnarray*}
\noindent With this choice of parametrization shifts in $t$ and
$\phi$ are realized linearly. Therefore this choice makes sure that all
fermions and the other bosons remain unchanged under the action of
group elements corresponding to shifts in $t$ and $\phi$, which is
a desirable feature when planning to impose a uniform light-cone
gauge \cite{alday-2008-089}.

\smallskip

The choice of the matrix $T_{34}+T_{56}$ might be unexpected, but
it is of course exactly this matrix that generates shifts in
$\phi$ on the full $\CP^3$ coset element. Further details on this
and the parametrization of $\CP^3$ employed here are discussed in
Appendix \ref{GandTmatrices}.

\subsection{Linearly realized bosonic symmetries}

The global bosonic symmetry group $\mbox{USP}(2,2)\times
\mbox{SO}(6)$ acts on the coset element by multiplication from the
left. For $G \in \mbox{USP}(2,2)\times \mbox{SO}(6)$ we have
\[
G g = g^\prime g_c,
\]
\noindent where $g_c$ is a compensating transformation from
$\mbox{SO}(3,1)\times \mbox{U}(3)$. Different coset
parametrizations have consequences as to which symmetries are
linearly realized, and on which fields they are linearly realized.
For example for the coset parametrization $g = g_\chi g_B$, the
fermions undergo an adjoint linear action by $G$.

\smallskip

While our choice of coset parametrization is well suited for
imposition of a uniform light-cone gauge, it does not allow for a
linear realization of all bosonic symmetries. Here we will
determine the symmetries that are linearly realized, which will
then form the manifest bosonic symmetry group of the light-cone
gauge fixed Lagrangian.

\smallskip

The linearly realized symmetries will be given by the subgroup of
$\mbox{USP}(2,2)\times \mbox{SO}(6)$ forming the centralizer of
group elements corresponding to the $\mbox{U}(1)$ isometries of
shifts in $t$ and $\phi$. The Lie algebra centralizer of these
$\alg{u}(1)$-isometries in $\alg{usp}(2,2) \oplus \alg{so}(6)$ is
easily found to be the subalgebra
\begin{equation}
\label{eq:algcentr} \mathfrak{C} = \alg{su}(2) \oplus \alg{su}(2)
\oplus \alg{u}(1),
\end{equation}
\noindent in agreement with
\cite{sundin-2008,Bykov:2009jy,Zarembo:2009au}, where this was
found for different (coset) parametrizations. Here the first
factor corresponds to $\alg{su}(2) \subset \alg{usp}(2,2)$, while
the second and third correspond to $\alg{su}(2) \oplus
\alg{u}(1)\subset \alg{so}(6)$. This stems from the fact that in
the AdS sector the centralizer is spanned by the $\Gamma^{ij}$
for $i,j=1,2,3$, since these all commute with $\Gamma^0$, forming
a four dimensional reducible representation of the Lie algebra
$\alg{su}(2)$, while in the $\mathbb{CP}$ sector it is spanned by
the set $\{T_{12}, (T_{45}-T_{36}),  (T_{34}-T_{56}),
(T_{35}+T_{46})\}$, commuting with $T_{34} + T_{56}$, forming a
reducible representation of the Lie algebra $\alg{su}(2) \oplus
\alg{u}(1)$.

\smallskip

Note that a group element, $G$, generated by elements of the
algebra, (\ref{eq:algcentr}), has the following action on the
coset element
\[
G g = \Lambda G g_\chi G^{-1} G g_B G^{-1} G,
\]
\noindent  since $G \Lambda G^{-1} = \Lambda$. As both the
centralizer subalgebras $\alg{su}(2)$ and $\alg{su}(2)\oplus
\alg{u}(1)$ are in fact subalgebras of the coset denominator Lie
algebras, we see that we can recognize the last $G$ above as
nothing else than a compensating transformation from
$\mbox{SO}(3,1)\times\mbox{U}(3)$. Thus we see that the element
$G$ has a linear adjoint action on the fermionic coset element
$g_\chi$ and the bosonic element $g_B$.

\smallskip

While the four dimensional reducible representation of
$\mbox{SU}(2)$ in the AdS sector is, by our convention for the
gamma matrices, in canonical direct sum form (see below), the set
of $T$-matrices above does not present a reducible representation
in such canonical form. By doing a change of basis that
diagonalizes $T_{12}$, $T_{34}$, and $T_{56}$, we obtain the more
pleasant canonical direct sum representation of our symmetry
group. Labelling the set of $T$-matrices introduced above as
$\{a_0,a_1,a_2,a_3\}$, we have
\begin{align*}
a_0 & \rightarrow \tilde{a}_0 = S a_0 S^{-1} = \mbox{diag$(i \sigma_3, 0, 0)$}\,, \\
a_i & \rightarrow \tilde{a}_i = \mbox{diag$(0, \frac{i
\sigma_i}{2}, \frac{i \sigma_i}{2})$}\, ,
\end{align*}
\noindent where the $\sigma_i$ are the Pauli matrices, and for
completeness
\begin{equation}
\label{coordtrans} S = \tfrac{1}{2} \left(
\begin{smallmatrix}
 i \sqrt{2} & \sqrt{2} & 0 & 0 & 0 & 0 \\
 -i \sqrt{2} & \sqrt{2} & 0 & 0 & 0 & 0 \\
 0 & 0 & -1 & i & i & 1 \\
 0 & 0 & -i & -1 & 1 & -i \\
 0 & 0 & 1 & i & -i & 1 \\
 0 & 0 & -i & 1 & 1 & i
\end{smallmatrix}
\right).
\end{equation}
\noindent In terms of two by two blocks, a group element, $G$, of
$\mbox{SU}(2)\times \mbox{SU}(2) \times \mbox{U}(1)$ in this basis
then takes the form
\begin{equation}
\label{eq:groupelem} G = \left(
\begin{array}{lllll}
 \alg{g}_1 & 0 & 0 & 0 & 0 \\
 0 & \alg{g}_1 & 0 & 0 & 0 \\
 0 & 0 & \tilde{\alg{g}}_{2} & 0 & 0 \\
 0 & 0 & 0 & \alg{g}_3 & 0 \\
 0 & 0 & 0 & 0 & \alg{g}_3
\end{array}
\right)\, .
\end{equation}
\noindent Here $\alg{g}_1$ is an element of the first
$\mbox{SU}(2)$, $\tilde{\alg{g}}_{2} =
\mbox{diag}(\alg{g}_2,\alg{g}_2^{-1})$ with $\alg{g}_2$ an element
of $\mbox{U}(1)$, and $\alg{g}_3$ is an element of the second
$\mbox{SU}(2)$.

\smallskip

To pick a $\kappa$-symmetry gauge choice that is manifestly
compatible with the above bosonic symmetries, let us first write
the fermionic element after the above change of basis and upon
imposition of the reality condition, $\tilde{\theta}$, as
\begin{equation}
\nonumber \tilde{\theta} = \frac{1}{2} \left(\begin{array}{lll}
\kappa_1 & \kappa_3 & \kappa_5 \\
\kappa_2 & \kappa_4 & \kappa_6 \\
\end{array}\right) ,
\end{equation}
\noindent where
\begin{align*}
\kappa_1 =   \left(
\begin{array}{ll}
 f_1 & f_2 \\
 f_3 & f_4
\end{array}
\right) = (\bar{\varkappa}_1,\varkappa_1), \mbox{\hspace{5pt}}&
\mbox{\hspace{5pt}} \kappa_2 =  \left(
\begin{array}{ll}
 -f_4^{*} & -f_3^{*} \\
 f_2^{*} & f_1^{*}
\end{array}
\right) = (\bar{\varkappa}_2,\varkappa_2)\\
\k_3 = \left(
\begin{array}{ll}
 f_5 & f_6 \\
 f_7 & f_8
\end{array}
\right), \mbox{\hspace{5pt}}& \mbox{\hspace{5pt}} \k_6 =  -i \sigma_2 \k_3^* \sigma_2 = \left(
\begin{array}{ll}
 -i f_8^{*} &  i f_7^{*} \\
  i f_6^{*} & -i f_5^{*}\end{array}
\right)\, ,
\end{align*}
\noindent and where we have labelled the fermions $f_i$. There are
similar relations between $\kappa_4$ and $\kappa_5$.

\smallskip

The specific $\kappa$-symmetry parameter, presented in Appendix
\ref{kappappendix}, allows us to remove eight fermions by removing
the blocks $\kappa_4$ and $\kappa_5$ from the fermionic element
$\tilde{\theta}$ giving
\begin{equation}
\label{eq:kgaugechoice} \tilde{\theta}_\kappa = \frac{1}{2}
\left(\begin{array}{lll}
\kappa_1 & \kappa_3 & 0 \\
\kappa_2 & 0 & \kappa_6 \\
\end{array}\right).
\end{equation}
\noindent This is a natural choice to make, and most importantly
is manifestly compatible with the above bosonic symmetries. We
thus consider this a convenient gauge choice and hence will fix it
as such.

\smallskip

When the above change of basis is applied to the bosonic coset
element it also induces a natural structure there, as seen for the
fermionic sector already. The bosonic structure in the
$\CP$-sector becomes
\begin{align*}
Y = \sum_{i=1}^{5} y_{i} T_{i} \rightarrow \tilde{Y} & =
\frac{1}{\sqrt{2}}\left(
\begin{array}{llll}
 0 & 0 & 0_{_{1\times2}} & -\tilde{y}^\dagger \\
 0 & 0 & -y^\dagger & 0_{_{1\times2}} \\
 0_{_{2\times1}} & y & 0_{_{2\times2}} & -i \sqrt{2} I_{_{2\times2}} y_5 \\
\tilde{y} & 0_{_{2\times1}} & -i \sqrt{2} I_{_{2\times2}} y_5 &
0_{_{2\times2}}
\end{array}
\right).\\
\end{align*}
\noindent For completeness, the unaffected bosonic \ads -sector is
of the form
\[
\sum_{i=1}^3 x_i \Gamma^i = \left(\begin{array}{ll} 0 & X\\-X & 0
\end{array}\right).
\]
\noindent In the above we have introduced the vectors and matrices
\begin{align*}
X = & X^\dagger = \left(
\begin{array}{cc}
 x_1 & x_2-i x_3 \\
 x_2+i x_3 & -x_1
\end{array}
\right)\\
y = & \left(
\begin{array}{l}
 i y_{1}+y_{2}+y_{3}-i y_{4} \\
 -y_{1}+i y_{2}-i y_{3}-y_{4}
\end{array}
\right)\\
\tilde{y} = & \left(
\begin{array}{l}
 i y_{1}-y_{2}+y_{3}+i y_{4} \\
 y_{1}+i y_{2}+i y_{3}-y_{4}
\end{array}
\right).
\end{align*}
\noindent This natural structure can now be used in the
construction of a manifestly covariant Lagrangian, but for clarity let us first explicitly give
the action of the manifest symmetry group on the matrices and
vectors just introduced. With the group element $G$ of the form
(\ref{eq:groupelem}), its action on the bosonic matrices defined
above is explicitly given by
\[
X \rightarrow \alg{g}_1 X \alg{g}_1^{-1}, \mbox{\hspace{10pt}} y
\rightarrow \alg{g}_3 y \alg{g}_2, \mbox{\hspace{10pt}} \tilde{y}
\rightarrow \alg{g}_3 y \alg{g}_2^{-1},
\]
\noindent while its action on the fermionic blocks on the other
hand is given by
\[
\bar{\varkappa}_{1,2}  \rightarrow \alg{g}_1 \bar{\varkappa}_{1,2}
\alg{g}_2^{-1}, \mbox{\hspace{10pt}} \varkappa_{1,2} \rightarrow
\alg{g}_1 \varkappa_{1,2} \alg{g}_2, \mbox{\hspace{10pt}}
\kappa_{3,6}  \rightarrow \alg{g}_1 \kappa_{3,6} \alg{g}_3^{-1}.
\]
\noindent With these it is easy to construct various covariant
couplings, which will be used in the next section.

\subsection{Manifestly covariant Lagrangian}

The explicit form of the Lagrangian, given by (\ref{lagrangianA2A2plusA1A3}), calculated to quartic order in the fields, is here cast into a manifestly covariant form using the above introduced block structure. For presentation purposes the Lagrangian is split into the purely bosonic sector and the fermionic sector, where the Wess-Zumino term is presented separately. Concretely,
\begin{equation}
\label{covlag}
\mathcal{L} = \gamma^{\a\b}(\mathcal{L}_{\mbox{\tiny b}} + \mathcal{L}_{\mbox{\tiny bf}} + \mathcal{L}_{\mbox{\tiny f}})_{\a\b} + \epsilon^{\a\b}(\mathcal{L}_{\mbox{\tiny bfWZ}} + \mathcal{L}_{\mbox{\tiny fWZ}})_{\a\b}
\end{equation}
Because of the special character of the $y_5$ field, as mentioned above and discussed in Appendix \ref{GandTmatrices}, it is expanded around $\tfrac{\pi}{4}$. Below we have written $\upsilon_5 = y_5 - \tfrac{\pi}{4}$.

\smallskip

To start with the purely bosonic Lagrangian, not expanded for sake of brevity, is given by
{\small
\begin{align*}
(\mathcal{L}_{\mbox{\tiny b}})_{\a\b} & = -\tfrac{1}{2} \left(\pa_\a t \pa_\b t - \pa_\a \phi \pa_\b \phi \right) -\tfrac{1}{2} \cosh{4 \rho} \pa_\a t \pa_\b t  \\
& -\frac{\left((\cos{4 \psi}+3) \psi ^4-8 (\cos{2 \psi}+2) \sin^2{\psi} y_5^2 \psi ^2+8 \sin^4{\psi} y_5^4\right) \pa_\a \phi \pa_\b \phi}{8 \psi^4}\\
& - \frac{i \sin^2{\psi} \left(\psi ^2 \cos^2{\psi}-\sin^2{\psi} y_5^2\right) \pa_\a \phi }{\psi ^4} \left( \partial_{\beta}y^{\dagger}y - h.c. \right)\\
& + \frac{\left(8 \rho^2 - \cosh (4 \rho)+1\right) \partial_{\alpha}\rho\partial_{\beta}\rho}{2 \rho ^2} + \frac{\sinh^2{2 \rho}}{2 \rho ^2} \text{Tr}(\partial_{\alpha}X\partial_{\beta}X) \\
& - \frac{4 \left(\left(\sin^2{\psi}-\psi^2 \right) \partial_{\alpha}\psi\partial_{\beta}\psi -\sin^2{\psi} \partial_{\alpha} y_{5} \partial_{\beta} y_{5} \right)}{\psi^2} \\
& + \frac{2 \sin^2{\psi}}{\psi ^2}\left( \partial_{\alpha}y^{\dagger} \partial_{\beta}y \right) + \frac{\sin^4{\psi}}{4 \psi^4}\left( \partial_{\alpha}y^{\dagger}y - h.c.\right)\left( \partial_{\beta}y^{\dagger}y - h.c. \right),
\end{align*}}
\noindent where we have introduced the invariant quantities $\rho = \sqrt{x_i x^i}$ and $\psi = \sqrt{y_j y^j}$, with $i=1,\ldots,3$ and $j=1,\ldots,5$. In the above, $h.c.$ denotes Hermitian conjugation of the term within the enclosing brackets; below where there are no brackets this hence stands for the Hermitian conjugate of the complete term. This bosonic Lagrangian can be compared to the one presented in \cite{sundin-2008,Kalousios:2009ey} where the bosonic string was studied upon imposition of a uniform light-cone gauge; the following parts of the action should be of more novel interest. The boson-fermion interactions in the kinetic term of the Lagrangian are at quartic order given by (continued on the next page)
{\small
\begin{align*}
(\mathcal{L}_{\mbox{\tiny bf}})_{\a\b}= & \tfrac{1}{2} \pa_\a t \pa_\b t \left(1 + 2  \mbox{Tr}\left(X^2\right)\right) \mbox{Tr}\left(\k_1^\dagger \k_1\right) - 2 i \pa_\a t \mbox{Tr}\left(\k_1^\dagger X \pa_\b X \k_1\right) \\
& + \tfrac{8 i}{\pi^2} \pa_\a t y^\dagger \pa_\b y (\bar{\varkappa}_1^\dagger \bar{\varkappa}_1 - \varkappa_1^\dagger \varkappa_1)\\
\end{align*}

\begin{align*}
& + \tfrac{1}{\sqrt{2}\pi^2}\pa_\a (2 t + \phi) \Big( (2\pi - 8 \upsilon_5)(\pa_\b \tilde{y}^\dagger \k_3^\dagger \bar{\varkappa}_1 - i \pa_\b y^\dagger \k_3^\dagger \varkappa_1) + 4 \pi (\upsilon_5 \pa_\b \tilde{y}^\dagger \k_3^\dagger \bar{\varkappa}_1 - i \pa_\b \upsilon_5 y^\dagger \k_3^\dagger \varkappa_1)\\
& \hspace{20pt} + (-8 \pa_\b \upsilon_5 + i(-\pi +4 \upsilon_5)\pa_\b \phi)(\tilde{y}^\dagger \k_3^\dagger \bar{\varkappa}_1 - i y^\dagger \k_3^\dagger \varkappa_1) + 2\pi i \upsilon_5 \pa_\b \phi (\tilde{y}^\dagger \k_3^\dagger \bar{\varkappa}_1 + 2 i y^\dagger \k_3^\dagger \varkappa_1)\Big)\\
& + \pa_\a (t +\phi) ( - 2 i \mbox{Tr}\left(\k_3^\dagger X \pa_\b X \k_3\right) + (\tfrac{1}{2} \pa_\b t (1+2\mbox{Tr}\left(X^2\right)) + (\tfrac{1}{2} - 2 \upsilon_5^2 - \tfrac{2}{\pi^2} y^\dagger y)\pa_\b \phi)\mbox{Tr}\left(\k_3^\dagger \k_3\right)\\
& \hspace{20pt}- \tfrac{4 i}{\pi^2} (\pa_\b y^\dagger \k_3^\dagger \k_3 y + \pa_\b \tilde{y}^\dagger \k_3^\dagger \k_3 \tilde{y})) - 2 \mbox{Tr}\left(\k_1^\dagger (X \pa_\a X - \pa_\a X X)\pa_\b \k_1\right)\\
& + i \pa_\a t (1 + 2\mbox{Tr}\left(X^2\right)) \mbox{Tr}\left(\pa_\b \k_1^\dagger \k_1 \right) - 2 \mbox{Tr}\left(\k_1^\dagger \pa_\a(t+i X) \pa_\b \k_2\right) \\
& + \tfrac{8}{\pi^2} \left(\pa_\a \varkappa_1^\dagger \varkappa_1 (y^\dagger \pa_\b y - \pa_\b y^\dagger y) + \pa_\a \bar{\varkappa}_1^\dagger \bar{\varkappa}_1 (\tilde{y}^\dagger \pa_\b \tilde{y} - \pa_\b \tilde{y}^\dagger \tilde{y})\right) \\
& + \tfrac{2\sqrt{2}i}{\pi^2}(\pi + 2(\pi-2) \upsilon_5)\pa_\a \tilde{y}^\dagger(\pa_\b \k_3^\dagger \bar{\varkappa}_1 - \k_3^\dagger \pa_\b  \bar{\varkappa}_1) + \tfrac{2\sqrt{2}}{\pi^2}(\pi - 4 \upsilon_5) \pa_\a y^\dagger (\pa_\b \k_3^\dagger \varkappa_1 - \k_3^\dagger \pa_\b \varkappa_1)\\
& + \tfrac{\sqrt{2}}{\pi^2}(-8 i \pa_\a \upsilon_5 + (\pi - 2(2+\pi) \upsilon_5)\pa_\a \phi)\tilde{y}^\dagger (\pa_\b \k_3^\dagger \bar{\varkappa}_1 - \k_3^\dagger \pa_\b  \bar{\varkappa}_1)\\
& + \tfrac{\sqrt{2}}{\pi^2} (4(\pi-2) \pa_\a \upsilon_5 - (i\pi + 4 (\pi-1) i \upsilon_5)\pa_\a \phi)y^\dagger (\pa_\b \k_3^\dagger \varkappa_1 - \k_3^\dagger \pa_\b \varkappa_1)\\
& + i ((1+2\mbox{Tr}\left(X^2\right)) \pa_\a t + (1-2 \upsilon_5^2 -\tfrac{4}{\pi^2} y^\dagger y) \pa_\a \phi)\mbox{Tr}\left(\pa_\b \k_3^\dagger \k_3\right) \\
& - 2 \mbox{Tr}\left(\k_3^\dagger(X\pa_\a X - \pa_\a X X)\pa_\b \k_3\right) +\tfrac{8}{\pi^2}(y^\dagger \pa_\a \k_3^\dagger \k_3 \pa_\b y - \pa_\a y^\dagger \pa_\b \k_3^\dagger \k_3 y) \\
& - \tfrac{4}{\pi^2}(y^\dagger \pa_\a y - \pa_\a y^\dagger y) \mbox{Tr}\left(\k_3^\dagger \pa_\b \k_3\right) + h.c.
\end{align*}}
\noindent Here we note that couplings containing $t$ or $\phi$ will upon imposition of a light-cone gauge reduce in order, the Lagrangian presented here contains all terms which afterwards will be of quartic order or lower. The rather unpleasant looking coefficients arise due to the expansion of $y_5$ around $\pi/4$, the terms from which they originate before expanding $y_5$ look quite natural. Continuing, the quartic fermionic couplings are given by (continued on the next page)

{\small
\begin{align*}
(\mathcal{L}_{\mbox{\tiny f}})_{\a\b} = & \tfrac{1}{4} \mbox{Tr}\Big(\pa_\a \k_2^\dagger \k_1 \pa_\b \k_2^\dagger \k_1 - \pa_\a \k_2^\dagger \pa_\b \k_1 \k_2^\dagger \k_1 + \pa_\a \k_1^\dagger \pa_\b \k_2 \k_2^\dagger \k_1 - \pa_\a \k_1^\dagger \k_1 \pa_\b \k_1^\dagger \k_1 \\
& \hspace{25pt}+ \pa_\a \k_1^\dagger \pa_\b \k_1 \k_1^\dagger \k_1 - \pa_\a \k_2^\dagger \pa_\b \k_2 \k_1^\dagger \k_1 + \k_2^\dagger \pa_\a \k_2 \k_1^\dagger \pa_\b \k_1 - \k_2^\dagger \pa_\a \k_1 \k_1^\dagger \pa_\b \k_2 \Big)\\
& + \tfrac{1}{4}\mbox{Tr}\Big( \k_1^\dagger \pa_\a \k_3 \pa_\b \k_3^\dagger \k_1 + \pa_\a \k_1^\dagger \k_3 \k_3^\dagger  \pa_\b \k_1 - 4 \pa_\a \k_1^\dagger \k_3 \pa_\b \k_3^\dagger \k_1 + 2 \pa_\a \k_1^\dagger \pa_\b \k_3 \k_3^\dagger \k_1 \\
& \hspace{30pt} + 2 \k_2^\dagger \k_3 \pa_\a \k_3^\dagger \pa_\b \k_2 - 2\k_2^\dagger \pa_\a \k_3  \k_3^\dagger \pa_\b \k_2 \Big) \\
& + \tfrac{i}{2} \left(\bar{\varkappa}_1^\dagger \pa_\a \k_3 \pa_\b \k_6^\dagger \bar{\varkappa}_2 + \pa_\a \bar{\varkappa}_1^\dagger  \k_3 \k_6^\dagger \pa_\b \bar{\varkappa}_2 - \pa_\a \bar{\varkappa}_1^\dagger  \k_3 \pa_\b \k_6^\dagger \bar{\varkappa}_2 - \bar{\varkappa}_1^\dagger \pa_\a \k_3 \k_6^\dagger  \pa_\b \bar{\varkappa}_2\right)\\
& + \tfrac{i}{6} \pa_\a t \Big(\bar{\varkappa}_1^\dagger \bar{\varkappa}_1 \bar{\varkappa}_1^\dagger \pa_\b \bar{\varkappa}_1 + \varkappa_1^\dagger \varkappa_1 \varkappa_1^\dagger \pa_\b \varkappa_1 + 4 (\bar{\varkappa}_1^\dagger \bar{\varkappa}_1 \varkappa_1^\dagger \pa_\b \varkappa_1 + \varkappa_1^\dagger \varkappa_1 \bar{\varkappa}_1^\dagger \pa_\b \bar{\varkappa}_1 ) \\
& \hspace{25pt} - \bar{\varkappa}_1^\dagger \varkappa_1\varkappa_1^\dagger \pa_\b \bar{\varkappa}_1 - \varkappa_1^\dagger \bar{\varkappa}_1 \bar{\varkappa}_1^\dagger \pa_\b \varkappa_1 + 8\mbox{Tr}\left(\pa_\b \k_1^\dagger \k_3 \k_3^\dagger \k_1 - \k_1^\dagger \pa_\b \k_3 \k_3^\dagger \k_1 \right)\\
&  \hspace{25pt} + 3 \mbox{Tr}\left(\k_2^\dagger \pa_\b \k_3 \k_3^\dagger \k_2 \right)\Big)\\
& + \tfrac{1}{4} \pa_\a (2 t + \phi) \left(\varkappa_1^\dagger \pa_\b \k_3 \k_6^\dagger \varkappa_2 + \varkappa_1^\dagger \k_3 \pa_\b \k_6^\dagger \varkappa_2 - \pa_\b \varkappa_1^\dagger \k_3 \k_6^\dagger \varkappa_2 - \varkappa_1^\dagger \k_3 \k_6^\dagger \pa_\b \varkappa_2 \right) \\
\end{align*}
\begin{align*}
& + i \pa_\a (t + \phi) \left(\tfrac{1}{2} \mbox{Tr}\left(\pa_\b \k_2^\dagger \k_3 \k_3^\dagger \k_2\right) + \tfrac{1}{3} \mbox{Tr}\left(\pa_\b \k_3^\dagger \k_3 \k_3^\dagger \k_3\right)\right) \\
& + i \pa_\a \phi\mbox{Tr}\left(\k_1^\dagger \k_3 \k_3^\dagger \pa_\b \k_1 + \tfrac{1}{3} \k_1^\dagger \pa_\b \k_3 \k_3^\dagger \k_1 \right)\\
& + \pa_\a t \pa_\b t \left(\tfrac{1}{12} \mbox{Tr}\left(2\k_1^\dagger \k_1 \k_1^\dagger \k_1 + \k_1^\dagger \k_1 \k_2^\dagger \k_2\right) + \tfrac{1}{8} \mbox{Tr}\left(\k_1^\dagger \k_1 \right) \mbox{Tr}\left(\k_2^\dagger \k_2 \right)\right)\\
& + \tfrac{1}{4}\pa_\a (2 t + \phi) \pa_\b (2 t + \phi) \left(\tfrac{i}{2} \bar{\varkappa}_2 \k_6 \k_3^\dagger \bar{\varkappa}_1 + \tfrac{7}{12} \mbox{Tr}\left(\k_1^\dagger \k_3 \k_3^\dagger \k_1\right) -\tfrac{1}{4} \mbox{Tr} \left(\k_1^\dagger \k_1\right) \mbox{Tr} \left(\k_3^\dagger \k_3\right) \right)\\
& + \pa_\a \phi \pa_\b \phi \left(-\tfrac{1}{24}\mbox{Tr}\left(\k_1^\dagger \k_3 \k_3^\dagger \k_1\right) + \tfrac{1}{16} \mbox{Tr}\left(\k_1^\dagger \k_1 \right) \mbox{Tr}\left(\k_3^\dagger \k_3 \right) \right)\\
& + \tfrac{1}{6} \pa_\a (t+\phi) \pa_\b (t+\phi) \mbox{Tr}\left(\k_3^\dagger \k_3 \k_3^\dagger \k_3\right) + h.c.
\end{align*}}
\noindent The asymmetry in these terms between $t$ and $\phi$ results from the different dimensionalities of the AdS and $\mathbb{CP}^3$ spaces; explicitly $t$ can couple to all fermions directly, whereas $\phi$ cannot. Continuing, the contributions from the Wess-Zumino term are divided in the same fashion as the above terms, first we have the boson-fermion interaction terms
{\small
\begin{align*}
(\mathcal{L}_{\tiny {bf}}^{\mbox{\tiny WZ}})_{\a\b}=&
\partial_{\alpha} t \partial_{\beta} \phi\tfrac{2 \sqrt{2}}{\pi
}\left( \bar{\varkappa}_1 ^{\dagger}X\kappa_3\tilde{y}-i
\varkappa_1 ^{\dagger}X\kappa_3y \right)
  - i \left( \text{Tr}\left(X^2\right)- \tilde{y}^{\dagger} \tilde{y}\right) \partial_{\beta}t\left( \partial_{\alpha} \bar{\varkappa}_1 ^{\dagger} \bar{\varkappa}_2-  \bar{\varkappa}_1 ^{\dagger} \partial_{\alpha} \bar{\varkappa}_2 \right) \nonumber\\
& +\tfrac{4 \sqrt{2}}{\pi}\left(\partial_{\beta}
t+\partial_{\beta} \phi\right)\left( \partial_{\alpha}
\bar{\varkappa}_1 ^{\dagger}X\kappa_3\tilde{y}+i\partial_{\alpha}
\varkappa_1 ^{\dagger}X\kappa_3y  \right)\nonumber
 + \tfrac{4 \sqrt{2}}{\pi}\partial_{\beta} t\left( \varkappa_1 ^{\dagger}X\partial_{\alpha} \kappa_3y+i \bar{\varkappa}_1 ^{\dagger}X\partial_{\alpha} \kappa_3\tilde{y} \right)\nonumber\\
& -\tfrac{1}{4}\left(1-2 \upsilon_5^2\right)\text{Tr}\left(
\kappa_6 ^{\dagger} \partial_{\alpha}\kappa_3
\right)\left(\partial_{\beta} t+\partial_{\beta}
\phi\right)\nonumber
  +\tfrac{1}{2} i  \left(  \bar{\varkappa}_1 ^{\dagger} \partial_{\alpha} \bar{\varkappa}_2-  \varkappa_1 ^{\dagger} \partial_{\alpha} \varkappa_2+  \bar{\varkappa}_2 ^{\dagger} \partial_{\alpha} \bar{\varkappa}_1
 -  \varkappa_2 ^{\dagger} \partial_{\alpha} \varkappa_1 \right) \partial_{\beta} t \nonumber\\
 &+2\upsilon_5\left(\partial_{\beta} t+\partial_{\beta} \phi\right) \text{Tr}\left( \partial_{\alpha} \kappa_3 ^{\dagger}X \kappa_3 +\partial_{\alpha} \kappa_6 ^{\dagger}X \kappa_6 \right) \nonumber\\
&-2 \left( \partial_{\alpha} \bar{\varkappa}_1 ^{\dagger}X
\bar{\varkappa}_1- \partial_{\alpha} \varkappa_1 ^{\dagger}X
\varkappa_1 \right) \partial_{\beta} t
  + \sqrt{2}i \left(\partial_{\beta} t+\partial_{\beta} \phi\right) \left(\partial_{\alpha} \varkappa_2 ^{\dagger} \kappa_3y  \right)
 -\sqrt{2}i \partial_{\beta} t \left( \varkappa_2 ^{\dagger} \partial_{\alpha} \kappa_3y \right)\nonumber\\
& - \left( \tfrac{2 \sqrt{2}}{\pi}+\tfrac{4 \sqrt{2}}{\pi^2}
\left(\pi-2\right)\upsilon_5\right) \left(\left(\partial_{\beta}
t+\partial_{\beta} \phi\right)\partial_{\alpha} \bar{\varkappa}_2
^{\dagger} \kappa_3\tilde{y}\nonumber
- \partial_{\beta} t \,\bar{\varkappa}_2 ^{\dagger} \partial_{\alpha} \kappa_3\tilde{y} \right)\nonumber\\
&  - i \left(1-2
\upsilon_5^2\right)\text{Tr}\left(\partial_{\alpha}
\kappa_6^{\dagger} \partial_{\beta} \kappa_3  \right)
 +\tfrac{1}{2} \left( \partial_{\alpha} \bar{\varkappa}_1 ^{\dagger} \partial_{\beta} \kappa_{2}- \partial_{\alpha} \varkappa_1 ^{\dagger} \partial_{\beta} \varkappa_2 \right)\nonumber
 -\tfrac{1}{2}\left( \partial_{\alpha} \bar{\varkappa}_2 ^{\dagger} \partial_{\beta} \kappa_{1}- \partial_{\alpha} \varkappa_2 ^{\dagger} \partial_{\beta} \varkappa_1 \right)\nonumber\\
& -4i \upsilon_5\text{Tr}\left( \partial_{\alpha} \kappa_3
^{\dagger}X \partial_{\beta} \kappa_3 \right)
 +2 i \left(\partial_{\alpha} \bar{\varkappa}_1 ^{\dagger}X \partial_{\beta} \bar{\varkappa}_1- \partial_{\alpha} \varkappa_1 ^{\dagger}X \partial_{\beta}\varkappa_1 \right)\nonumber\\
 & + \tfrac{4 \sqrt{2}}{\pi^2}\left(\pi-4 \upsilon_5\right)\left(\partial_{\alpha} \bar{\varkappa}_1 ^{\dagger} \partial_{\beta} \kappa_6\tilde{y} \right)
 +i  \tfrac{4\sqrt{2}}{\pi^2}\left(\pi +2\left(\pi -2\right) \upsilon_5 \right)\left(\partial_{\alpha} \bar{\varkappa}_1 ^{\dagger} \partial_{\beta} \kappa_6\tilde{y} \right)+h.c.\nonumber
 \end{align*}
\noindent As the final piece, the fermionic quartic terms are
given by
{\small
\begin{align*}
(\mathcal{L}_{\tiny f}^{\mbox{\tiny WZ}})_{\a\b}= &- \tfrac{1}{8}
\partial_{\alpha} t \partial_{\beta}\phi \left(\left(\varkappa_1
^{\dagger} \kappa_6 \kappa_6 ^{\dagger} \varkappa_2-
\bar{\varkappa}_1 ^{\dagger} \kappa_6 \kappa_6 ^{\dagger}
\bar{\varkappa}_2\right)
+ i\left(   \bar{\varkappa}_1 ^{\dagger}  \kappa_3   \kappa_6 ^{\dagger} \bar{\varkappa}_1+   \varkappa_1 ^{\dagger} \kappa_3 \kappa_6 ^{\dagger} \varkappa_1\right)\right) \nonumber\\
& +\tfrac{5}{6} i\left(  \varkappa_2 ^{\dagger} \varkappa_1
\varkappa_2 ^{\dagger} \partial_{\alpha} \varkappa_2-
\bar{\varkappa}_2 ^{\dagger} \bar{\varkappa}_1 \bar{\varkappa}_2
^{\dagger} \partial_{\alpha} \kappa_{2}\right) \partial_{\beta} t
+\tfrac{1}{6} i \left(  \varkappa_2 ^{\dagger} \partial_{\alpha} \varkappa_1    \varkappa _1 ^{\dagger}  \varkappa _1 -\bar{\varkappa}_2 ^{\dagger} \partial_{\alpha} \bar{\varkappa}_1    \bar{\varkappa} _1 ^{\dagger}  \bar{\varkappa} _1  \right)\partial_{\beta} t\nonumber\\
& +\tfrac{7}{6} i  \left( \partial_{\alpha} \bar{\varkappa}_2
^{\dagger} \bar{\varkappa}_1 \varkappa_2 ^{\dagger} \varkappa_2
 - \partial_{\alpha} \varkappa_2 ^{\dagger} \varkappa_1 \bar{\varkappa}_2 ^{\dagger} \bar{\varkappa}_{2}\right) \partial_{\beta} t  +\tfrac{1}{6}\left( {\varkappa}_2 ^{\dagger}   \kappa_3  \partial_{\alpha}\kappa_6 ^{\dagger}  {\varkappa}_2+   \bar{\varkappa}_2 ^{\dagger}  \kappa_3 \partial_{\alpha} \kappa_6 ^{\dagger} \bar{\varkappa}_2 \right)\left(  \partial_{\beta} t-  \partial_{\beta} \phi\right)\nonumber\\
 &  -\tfrac{1}{3} i \left(   \bar{\varkappa}_2 ^{\dagger} \partial_{\alpha} \kappa_6   \kappa_6 ^{\dagger} \bar{\varkappa}_1-   \varkappa_2 ^{\dagger} \partial_{\alpha} \kappa_6 \kappa_6 ^{\dagger} \varkappa_1\right) \partial_{\beta} t
 +\tfrac{1}{6} i   \left(   \bar{\varkappa}_2 ^{\dagger} \partial_{\alpha} \kappa_3   \kappa_3 ^{\dagger} \bar{\varkappa}_1
 -   \varkappa_2 ^{\dagger} \partial_{\alpha} \kappa_3 \kappa_3 ^{\dagger} \varkappa_1\right) \partial_{\beta} t\nonumber\\
&+  \tfrac{1}{12}\left(    {\varkappa}_2 ^{\dagger}
\partial_{\alpha} \kappa_3  \kappa_6 ^{\dagger}  {\varkappa}_1
 -   \bar{\varkappa}_2 ^{\dagger}  \partial_{\alpha}\kappa_3  \kappa_6 ^{\dagger} \bar{\varkappa}_2\right) \left(8\partial_{\beta} t-5\partial_{\beta} t \right)\nonumber\\
&+ \tfrac{1}{6} \left(\partial_{\beta} t+\partial_{\beta}
\phi\right)\left( 2\left(  \partial_{\alpha} \bar{\varkappa}_1
^{\dagger} \kappa_3   \kappa_6 ^{\dagger} \bar{\varkappa}_1
 +  \partial_{\alpha} \varkappa_1 ^{\dagger} \kappa_3\kappa_6 ^{\dagger} \varkappa_1\right) \nonumber
 - \left(  \partial_{\alpha} \bar{\varkappa}_1 ^{\dagger} \kappa_6   \kappa_3 ^{\dagger} \bar{\varkappa}_1
 +  \partial_{\alpha} \varkappa_1 ^{\dagger} \kappa_6 \kappa_3 ^{\dagger} \varkappa_1\right)   \right)\nonumber\\
& +\tfrac{1}{6} i \left(   \bar{\varkappa}_2 ^{\dagger} \kappa_3
\kappa_3 ^{\dagger} \partial_{\alpha} \bar{\varkappa}_1
 -   \varkappa_2 ^{\dagger} \kappa_3 \kappa_3 ^{\dagger} \partial_{\alpha} \varkappa_1\right) \partial_{\beta} t\nonumber
 -i \left(  \partial_{\alpha} \bar{\varkappa}_2 ^{\dagger} \kappa_3   \kappa_3 ^{\dagger} \bar{\varkappa}_1
 -  \partial_{\alpha} \varkappa_2 ^{\dagger} \kappa_3 \kappa_3 ^{\dagger} \varkappa_1\right) \left(\tfrac{8}{3} \partial_{\beta} t+ \partial_{\beta} \phi\right)\nonumber
 \end{align*}
\begin{align*}
&-\tfrac{1}{6} (  \partial_{\alpha} \bar{\varkappa}_{1} ^{\dagger}
\kappa_6   \kappa_3 ^{\dagger} \bar{\varkappa}_{1}
 +  \partial_{\alpha} \varkappa_{1} ^{\dagger} \kappa_6 \kappa_3 ^{\dagger} \varkappa_{1})  (\partial_{\beta} t-\partial_{\beta} \phi)
 -\tfrac{1}{6} \left( \partial_{\alpha} \bar{\varkappa}_1 ^{\dagger} \partial_{\beta} \bar{\varkappa}_1 \varkappa_2 ^{\dagger} \varkappa_1-\partial_{\alpha} \varkappa_1 ^{\dagger} \partial_{\beta} \varkappa_1  \bar{\varkappa}_2 ^{\dagger} \bar{\varkappa}_1\right)\nonumber\\
& + \tfrac{1}{6} \left(\partial_{\beta} t+\partial_{\beta} \phi\right)\text{Tr}\left(2\kappa_6 ^{\dagger} \kappa_3  \kappa_6 ^{\dagger} \partial_{\alpha} \kappa_6+\kappa_6 ^{\dagger} \partial_{\alpha} \kappa_3  \kappa_3 ^{\dagger} \kappa_3 +    \kappa_6 ^{\dagger} \kappa_3  \kappa_3 ^{\dagger} \partial_{\alpha} \kappa_3 \right) \nonumber\\
 &  -\tfrac{1}{3}\left( \partial_{\alpha} \bar{\varkappa}_1 ^{\dagger} \kappa_{1}  \partial_{\beta} \bar{\varkappa}_2 ^{\dagger} \bar{\varkappa}_1+ \partial_{\alpha} \bar{\varkappa}_1 ^{\dagger} \bar{\varkappa}_1\partial_{\beta} \varkappa_2 ^{\dagger} \varkappa_1
   - \partial_{\alpha} \varkappa_1 ^{\dagger} \varkappa_1  \partial_{\beta} \bar{\varkappa}_2 ^{\dagger} \bar{\varkappa}_1-\partial_{\alpha} \varkappa_1 ^{\dagger} \varkappa_1\partial_{\beta} \varkappa_2 ^{\dagger} \varkappa_1\right)\nonumber\\
&  +\tfrac{1}{3} \left(\partial_{\alpha} \bar{\varkappa}_2
^{\dagger} \bar{\varkappa}_2  \partial_{\beta} \bar{\varkappa}_2
^{\dagger} \bar{\varkappa}_1
 + \partial_{\alpha} \bar{\varkappa}_2 ^{\dagger} \bar{\varkappa}_2\partial_{\beta} \varkappa_2 ^{\dagger} \varkappa_1
 -\partial_{\alpha} \varkappa_2 ^{\dagger} \varkappa_2  \partial_{\beta} \bar{\varkappa}_2 ^{\dagger} \bar{\varkappa}_1
 - \partial_{\alpha} \varkappa_2 ^{\dagger} \varkappa_2  \partial_{\beta} \varkappa_2 ^{\dagger} \varkappa_1\right)\nonumber\\
 &+\tfrac{1}{2} \left( \partial_{\alpha} \varkappa_1 ^{\dagger} \partial_{\beta} \varkappa_2 \varkappa_2 ^{\dagger} \varkappa_2+\partial_{\alpha} \varkappa_1 ^{\dagger} \partial_{\beta} \varkappa_2 \bar{\varkappa}_2 ^{\dagger} \bar{\varkappa}_2\right)
 +\tfrac{1}{6} i   \text{Tr}\left(   \kappa_{1} ^{\dagger} \kappa_{1}  \partial_{\alpha} \kappa_6 ^{\dagger} \partial_{\beta} \kappa_3 \right)\nonumber\\
 &+\tfrac{1}{3} i \text{Tr}\left( \partial_{\alpha} \kappa_6 ^{\dagger} \kappa_{1}  \partial_{\beta} \kappa_{1} ^{\dagger} \kappa_3\right)
+  \tfrac{1}{3} \left(  \partial_{\alpha} \bar{\varkappa}_2 ^{\dagger} \kappa_3  \partial_{\beta} \kappa_3 ^{\dagger} \bar{\varkappa}_1-  \partial_{\alpha} \varkappa_2 ^{\dagger} \kappa_3\partial_{\beta} \kappa_3 ^{\dagger} \varkappa_1\right)\nonumber\\
&- \tfrac{1}{6}    \left( \partial_{\alpha} \bar{\varkappa}_2
^{\dagger} \partial_{\beta} \kappa_3   \kappa_3 ^{\dagger}
\bar{\varkappa}_1
 -   \partial_{\alpha} \varkappa_2 ^{\dagger} \partial_{\beta} \kappa_3   \kappa_3 ^{\dagger} \varkappa_1 \right)
 -\tfrac{1}{6} i   \left(\partial_{\alpha} \bar{\varkappa}_1 ^{\dagger} \partial_{\beta} \kappa_6   \kappa_3 ^{\dagger} \bar{\varkappa}_1  +  \partial_{\alpha} \varkappa_1 ^{\dagger} \partial_{\beta} \kappa_6 \kappa_3 ^{\dagger} \varkappa_1\right)\nonumber\\
&  -\tfrac{1}{6}\left(  \partial_{\alpha} \varkappa_2 ^{\dagger}
\kappa_6   \kappa_6 ^{\dagger} \partial_{\beta} \varkappa_1
-\partial_{\alpha} \bar{\varkappa}_2 ^{\dagger} \kappa_6
\kappa_6 ^{\dagger} \partial_{\beta} \bar{\varkappa}_1 \right)
 +\tfrac{1}{3} i \text{Tr}\left(\kappa_3 ^{\dagger} \partial_{\alpha} \kappa_3  \kappa_3 ^{\dagger} \partial_{\beta} \kappa_6 \right) +h.c.\nonumber
 \end{align*}
}
These terms combined presents the quartic Lagrangian in a manifest covariant form with respect to the bosonic light-cone symmetries. In the next section we will discuss an integrable reduction of the full model, and we would like to note here that the resulting Lagrangian presented below can almost directly be read of from the above upon taking the reduction into account.

\section{Integrable reductions for strings on ${\rm AdS}_4\times \mathbb{CP}^3$}

The full model given by the Lagrangian (\ref{lagrangianA2A2plusA1A3}) above, admits a consistent truncation, which upon imposition of the uniform light-cone gauge yields a purely fermionic Lagrangian containing two complex fermions. Being a consistent truncation, this model inherits the integrable structure possessed by the full Lagrangian. A remarkable and interesting fact is that this system is identical to the integrable system found in \cite{Alday:2005jm} for the $\AdS$ background.

\smallskip

We would like to truncate down to purely light-cone bosons, which will subsequently be removed by utilizing the reparametrization invariance of the action. The question then becomes which set of fermions admit such a truncation, but of course the remaining (light-cone) bosonic symmetry group gives a natural structure here. Given the representation of this group, it is again only natural to introduce the following block structure on a generic fermionic block $\eta$
\begin{equation*}
\eta = \tfrac{1}{2} \left(\begin{array}{lll}
\vartheta_1 & \vartheta_3 & \vartheta_5 \\
\vartheta_2 & \vartheta_4 & \vartheta_6 \\
\end{array}\right).
\end{equation*}
In order to truncate down to only light-cone bosons, i.e. the fields $t$ and $\phi$, we switch of all fermions besides either $\vartheta_3$ and $\vartheta_6$ or $\vartheta_4$ and $\vartheta_5$. Choosing to switch of $\vartheta_4$ and $\vartheta_5$ we are left with the block
\begin{equation*}
\left(\begin{array}{llllll}
0 & 0 & \eta^{31} & \eta^{33} & 0 & 0 \\
0 & 0 & \eta^{32} & \eta^{34} & 0 & 0 \\
0 & 0 & 0 & 0 & \eta^{61} & \eta^{63}\\
0 & 0 & 0 & 0 & \eta^{62} & \eta^{64}
\end{array}\right).
\end{equation*}
This leaves a truncated model with four complex fermionic degrees of freedom\footnote{The reality condition on the fermions relates those in $\vartheta_6$ to the ones in $\vartheta_3$.}. This model admits a further truncation, down to two complex fermions, achieved by switching off either the diagonal or off-diagonal terms in $\vartheta_3$ and $\vartheta_6$. Choosing here to set $\eta^{32}=\eta^{33}=\eta^{62}=\eta^{63}=0$, we obtain the smallest truncated model possible in this setting. Further truncations would lead to nonlinear constraints, due to couplings in the Wess-Zumino term of the Lagrangian schematically of the form $(t+\phi)\eta^{31}\eta^{34}$, with possible derivatives. Switching either of these fermionic fields off independently would lead to nonlinear constraints on the remaining field and the combination $t +\phi$ through the equation of motion of the removed fermion.

\smallskip

The reduced model thus obtained is manifestly integrable, being the theory of a free two dimensional massive Dirac fermion. It will nonetheless still be interesting to see how the Lax connection for the full model reduces for this specific model as shown further below.

\subsection{The Lagrangian of the reduced model}

The Lagrangian for the truncated model above gives a quadratic action for two complex fermions, equivalent to the integrable model presented in \cite{Alday:2005jm} found for the $\AdS$ background. Given our different choice of coset parametrization however, the fermions are uncharged under the ${\rm U}(1)$ shifts corresponding to shifts in $t$ and $\phi$ from the start, doing away with the need for a field redefinition of the fermions as done in \cite{Alday:2005jm}.

\smallskip

After substituting the truncation into the Lagrangian (\ref{lagrangianA2A2plusA1A3}) we find the following;
\begin{align*}
\mathcal{L}= & \tfrac{\sqrt{\lambda}}{4\pi} \big( \gamma^{\a\b} \left(\pa_\a t\pa_\b t -\pa_\a \phi \pa_\b \phi + \tfrac{i}{2} \pa_\a (t+\phi) \zeta_\b - \tfrac{1}{2} \pa_\a (t+\phi)\pa_\b (t+\phi) \Lambda \right) \\
& \nonumber - \tfrac{\k}{2} \epsilon^{\a\b} \pa_\a (t+\phi) \Omega_\b \big),
\end{align*}
\noindent where we have introduced the following abbreviations for the fermionic contributions
\begin{align*}
\zeta_\a & = 2 \left(\eta_{i} \pa_\a\eta^{i} + \eta^{i}\pa_\a\eta_{i}  \right),\\
\Lambda & = 2 \eta_{j}\eta^{j},\\
\Omega_\b & = -2 i \left( \eta^{31} \pa_\b \eta^{34} + \eta^{34} \pa_\b \eta^{31} +\eta_{31} \pa_\b \eta_{34} + \eta_{34} \pa_\b \eta_{31} \right).
\end{align*}
\noindent Here the reality condition has been imposed, we write $\eta^{i*} = \eta_{i}$, and the summations are of course over the relevant two fermions. The factor of $\sqrt{\lambda}/4\pi$ has been included here for comparison purposes. Upon rescaling our fermions as
\[
\eta^{31,34} \rightarrow \frac{e^{-i \frac{\pi}{4}}}{\sqrt{2}} \eta^{31,34} ,
\]
\noindent it should be clear that this action is identical to the one presented in \cite{Alday:2005jm}, where we note that there the string tension is taken to be $\tfrac{1}{2\pi\alpha '}=1$. This model simplifies to the greatest extent upon imposition of a uniform light-cone gauge, which utilizes the reparametrization invariance of the string action by introducing the following light-cone coordinates
\[
x_{+} = \tfrac{1}{2}(\phi+t), \hspace{10pt} x_{-} = \tfrac{1}{2}(\phi-t), \hspace{10pt} p_{+} = p_\phi + p_t, \hspace{10pt} p_{-} = p_t - p_\phi,
\]
\noindent and subsequently imposing the gauge choice
\[
x_{+} = \tau +\tfrac{m}{2} \sigma, \hspace{10pt} p_{+} = \mbox{constant} = P_{+} = E + J,
\]
\noindent where $m$ is the winding number which arises due to periodicity of the $\phi$ field. $E$ and $J$ are the Noether charges associated with the $U(1)$ isometries of
shifts in $t$ and $\phi$ respectively. This gauge fixing was carefully done in \cite{Arutyunov:2005hd} for this Lagrangian, hence the reader is referred there for the exact
details of the procedure. The upshot is the gauge fixed Lagrangian
\begin{equation}
\label{eq:lcgfL}
\mathcal{L} = -\frac{i}{4} P_+ \zeta_\tau + \frac{1}{2} \k m \Omega_\tau - \k \Omega_\sigma + \frac{1}{2} P_+ \Lambda.
\end{equation}
\subsection{Lax representation}

The integrability of classical superstrings on ${\rm AdS}_4\times \mathbb{CP}^3$ has been demonstrated in \cite{arutyunov-2008}; by construction there is in fact no essential difference from the $\AdS$ coset model. In the above we have shown that the ${\rm AdS}_4\times \mathbb{CP}^3$ model allows a reduction to a fermionic Lagrangian containing two complex fermions, yielding a manifestly integrable system of a free two dimensional massive Dirac fermion for the suitable choice of fixing a uniform light-cone gauge. Despite its manifest integrability, it will still be insightful to see how the general Lax connection reduces in the above truncation and uniform light-cone gauge.

\smallskip

The general Lax pair for this model, is a pair of ten by ten matrices, with zero curvature
\begin{equation}
\label{eq:curv}
\pa_\sigma {\rm L}_\tau - \pa_\tau {\rm L}_\sigma-\left[{\rm L}_\sigma,{\rm L}_\tau\right] = 0,
\end{equation}
\noindent following from the equations of motion and vice versa. For the reduced model, the appropriate Lax connection can be formulated in terms of two by two matrices. To connect with the notation in \cite{Arutyunov:2005hd}, we have as concise notation for the fermions
\[
\psi_1 = \eta_{31}, \hspace{5pt} \psi_2 = \eta^{34}, \hspace{5pt}  \psi_1^* = \eta^{31}, \hspace{5pt}  \psi_2^* = \eta_{34},
\]
\noindent and we introduce the even quantities
\begin{align*}
\varsigma_\a & = \psi_1 \pa_\a \psi_1^* + \psi_1^* \pa_\a \psi_1 - \psi_2 \pa_\a \psi_2^* - \psi_2^* \pa_\a \psi_2\\
\xi & = \psi_1 \psi_1^* + \psi_2 \psi_2^*.
\end{align*}
In terms of these quantities the components of the two by two Lax connection are given as
\begin{align*}
{\rm L}_\tau & =
\left(
\begin{array}{ll}
\tfrac{i (1+z^2)}{2(-1+z^2)}+ \tfrac{1}{8} \varsigma_\tau + \tfrac{i}{4} \xi & -\frac{e^{i\tfrac{\pi}{4}}(z(\pa_\tau \psi_2^* - i \psi_2^*)+(-i \pa_\tau \psi_1^* + \psi_1^*))}{\sqrt{2-2z^2}}\\ -\frac{e^{i\tfrac{\pi}{4}}(z(-i\pa_\tau \psi_2 + \psi_2)+(\pa_\tau \psi_1 - i \psi_1))}{\sqrt{2-2z^2}} & - \tfrac{i (1+z^2)}{2(-1+z^2)} + \tfrac{1}{8} \varsigma_\tau + \tfrac{i}{4} \xi
\end{array}
\right), \\
{\rm L}_\sigma & =
\left(
\begin{array}{ll}
\tfrac{i (z P_{+} + m + m z^2)}{4(-1+z^2)}+ \tfrac{1}{8} \varsigma_\sigma + \tfrac{i m}{8} \xi & -\frac{e^{i\tfrac{\pi}{4}}(z(2 \pa_\sigma \psi_2^* - i m \psi_2^*)+(-2i \pa_\sigma \psi_1^* + m \psi_1^*))}{2\sqrt{(2-2z^2)}}\\ -\frac{e^{i\tfrac{\pi}{4}}(z(-2i\pa_\sigma \psi_2 + m \psi_2)+(2 \pa_\sigma \psi_1 - i m \psi_1))}{2\sqrt{(2-2z^2)}} & - \tfrac{i (z P_{+} + m + m z^2)}{4(-1+z^2)}+ \tfrac{1}{8} \varsigma_\sigma + \tfrac{i m}{8} \xi
\end{array}
\right),
\end{align*}
\noindent where $z$ is the spectral parameter associated with this Lax pair. This reduced Lax connection contains the essential features of the original ten by ten Lax connection as of course it should; it is constructed directly from it by considering its independent entries and suitably combining them to obtain this equivalent two by two connection. It is easily checked that this Lax connection has zero curvature (\ref{eq:curv}) on shell, with the equations of motion following from (\ref{eq:lcgfL}) being
\begin{align*}
\pa_\tau \psi_1 & = \frac{-2 P_+ \left(m \psi_2 -2 i \pa_\sigma \psi_2 \right)+8 m \pa_\sigma \psi_1 -i P_+^2 \psi_1}{4 m^2-P_+^2}, \\
\pa_\tau \psi_2 & = \frac{-2 P_+ \left(m \psi_1 +2 i \pa_\sigma \psi_1 \right) + 8 m \pa_\sigma \psi_2 +i P_+^2 \psi_2 }{4 m^2-P_+^2}.
\end{align*}

\subsection{Other truncated models}

There are some good indications that other integrable truncations exist, to a sector containing one bosonic field from the AdS space and two complex fermions on top of the light-cone bosons $t$ and $\phi$. This would concretely for example be a truncation down to the bosonic fields $t$, $\phi$ and $x_1$, and the fermions $f_1$ and $f_4$, giving the fermionic block the structure
\[
\left(\begin{array}{llllll}
f_1 & 0 & 0 & 0 & 0 & 0 \\
0 & f_4 & 0 & 0 & 0 & 0 \\
-f_4^* & 0 & 0 & 0 & 0 & 0\\
0 & f_1^* & 0 & 0 & 0 & 0
\end{array}\right).
\]
\noindent In this truncated model obviously the field $\phi$ does not couple to the other fields and hence could also be removed, leaving just the AdS sector coupling to two complex fermions. This model has a Lagrangian nonlinear in the $x_1$ field, which extends till quartic order in the fermionic fields, leading to a considerably less simple Lagrangian than the one obtained above. This makes an explicit check of the integrability more involved and less insightful, hence we present it here as a possibility without explicit proof. Finally, further possible truncations would seem to require inclusion of even more fields and would hence be considerably less concise.

\section{Conclusions}

In this paper we have focussed on the  fermionic structure of the
coset sigma model describing Green-Schwarz superstrings on ${\rm
AdS}_4 \times \CP^3$. This model is classically integrable,
however, the question of quantum integrability is considerably
more involved and calls for further investigation. With a choice
of coset parametrization suitable for imposition of a light-cone
gauge, the manifest global bosonic symmetry algebra  of the model
was found to be $\mathfrak{C} = \alg{su}(2) \oplus \alg{su}(2)
\oplus \alg{u}(1)$, and $\k$-symmetry gauge fixing compatible
with this global symmetry was performed. Furthermore, with this
knowledge, the quartic Lagrangian of the sigma model has been cast
in a covariant form with respect to $\mathfrak{C}$, a result which
can be further used to compute the light-cone two-body S-matrix.
Analyzing the structure of the latter should present the first
step towards understanding quantum integrability of the model, by
allowing one to check the factorization property of the S-matrix.

\smallskip

In addition, by exploiting the classical integrable structure of
the full coset model, in this paper we have analyzed the question
whether there exist integrable truncations of the superstring
Lagrangian. We found that this question has an affirmative answer
by explicitly providing the (smallest) integrable fermionic model
which arises upon consistent truncation of the full coset model.
This truncation is again based in part on the manifest symmetry
algebra, and yields a model containing two complex fermions and
the bosonic light-cone fields $x_{\pm}$. Perhaps, the most
interesting fact about this model is that it is exactly the same
as the one arising from consistent truncation of the coset model
describing ${\rm AdS}_5\times {\rm S}^5$ superstrings. The reduced
model simplifies drastically upon imposition of the uniform
light-cone gauge where it coincides with a {\it free} model of two
complex fermions. In the ${\rm AdS}_5$/SYM duality this truncated
model is directly related to a certain closed sector of the dual
gauge theory. As such, given the exact agreement between the two
truncated models and the similarities between the two models in
general, it would be interesting to investigate whether this is
also the case for the current duality.

\section{Acknowledgements}

We would like to thank Gleb Arutyunov for valuable discussions.

\appendix

\section{The coset element and linear isometries}\label{GandTmatrices}

For reader's convenience, in this appendix we collect the matrices used in the construction of the coset element. The AdS$_4$ is parametrized by the following set of $\Gamma$ matrices

\begin{eqnarray}
 \nonumber
\begin{aligned} \Gamma^0&=&{\scriptsize \left(\begin{array}{cccc}
1 & 0 & 0 & 0 \\
0 & 1 & 0 & 0 \\
0 & 0 & -1 & 0 \\
0 & 0 & 0 & -1 \\
\end{array}\right)
}\, , ~~~~~ \Gamma^1={\scriptsize \left(\begin{array}{cccc}
0 & 0 & 1 & 0 \\
0 & 0 & 0 & -1 \\
-1 & 0 & 0 & 0 \\
0 & 1 & 0 & 0 \\
\end{array}\right)
}\, , \\
\nonumber \Gamma^2&=&{\scriptsize \left(\begin{array}{cccc}
0 & 0 & 0 & 1 \\
0 & 0 & 1 & 0 \\
0 & -1 & 0 & 0 \\
-1 & 0 & 0 & 0 \\
\end{array}\right)
}\, , ~~~~~ \Gamma^3={\scriptsize \left(\begin{array}{cccc}
0 & 0 & 0 & -i \\
0 & 0 & i & 0 \\
0 & i & 0 & 0 \\
-i & 0 & 0 & 0 \\
\end{array}\right)
}\, , \end{aligned}
\end{eqnarray}
such that these matrices satisfy the Clifford algebra $\{\Gamma^{\mu},\Gamma^{\nu}\}=2\eta^{\mu\nu}$, where $\eta^{\mu\nu}$ is Minkowski metric with signature $(1,-1,-1,-1)$. We also define $\Gamma^5=-i\Gamma^0\Gamma^1\Gamma^2\Gamma^3$ with the property $(\Gamma^5)^2={\mathbb I}$.

\smallskip

The charge conjugation matrix $C_4$ obeys $(\Gamma^{\mu})^t=-C_4\Gamma^{\mu} C_4^{-1}$ and in the present case it can be chosen as
\begin{eqnarray}
\nonumber C_4=i\Gamma^0\Gamma^3={\scriptsize \left(\begin{array}{cccc}
0 & 0 & 0 & 1 \\
0 & 0 & -1 & 0 \\
0 & 1 & 0 & 0 \\
-1 & 0 & 0 & 0 \\
\end{array}\right)
}\, .
\end{eqnarray}
Moreover we also require $(\Gamma^{\mu})^t=K_4\Gamma^{\mu} K_4^{-1}$, where $K_4=i I_2 \otimes \sigma_2$ and where $I_2$ is the $2 \times 2$ identity matrix. Lastly we define $\Gamma^{\mu\nu}\equiv\frac{1}{4}[\Gamma^{\mu},\Gamma^{\nu}]$.

\smallskip

To parametrize the $\CP^3$ coset space, we need to consider the space orthogonal to $\alg{u}(3)$ in $\alg{so}(6)$, which is spanned
by solutions to the following equation
\begin{eqnarray}\label{pCP}
\{K_6, Y\}=0\, ,
\end{eqnarray}
 The general solution to eq.(\ref{pCP}) is
six-parametric and is represented by a matrix
\begin{eqnarray}
\nonumber Y=  y_i T_i\, .
\end{eqnarray}
where we have introduced the six matrices $T_i$ which are Lie
algebra generators of $\alg{so}(6)$ along the $\CP^3$ directions:
\begin{eqnarray}
\nonumber \begin{aligned}
T_1&=E_{13}-E_{31}-E_{24}+E_{42}\, ,
~~~~~~~~T_2=E_{14}-E_{41}+E_{23}-E_{32}\, , \\
T_3&=E_{15}-E_{51}-E_{26}+E_{62}\, ,
~~~~~~~~T_4=E_{16}-E_{61}+E_{25}-E_{52}\, , \\
T_5&=E_{35}-E_{53}-E_{46}+E_{64}\, ,
~~~~~~~~T_6=E_{36}-E_{63}+E_{45}-E_{54}\, ,
\end{aligned}
 \end{eqnarray}
where $E_{ij}$ are the standard matrix unities, and where the $T_i$ matrices are normalized as${\rm tr}(T_iT_j)=-4\delta_{ij}$.
Moreover, the complete set of generators of $\alg{so}(6)$ are given by $T_{ij}=E_{ij}-E_{ji}$.

\smallskip

 In section \ref{cosetsymetriesandkappa}   we indicated that it would be desirable to extract the coordinates $t$ and $\phi$ from the general bosonic element, in anticipation of imposing a light-cone gauge choice on these coordinates. This was done by introducing the parametrization of the coset element as given by eq. (\ref{eq:cospar}). Here we will illustrate why this form is the appropriate choice.

\smallskip

$\CP^3$ can be parametrized in terms of the coset element
\begin{equation}
\label{eq:cpcospar}
g_\CP = e^Y,
\end{equation}
\noindent where
\[
Y = \sum_{i=1}^6 y_i T_i \, .
\]
We parametrize $\CP^3$ by introducing spherical coordinates, $(r,\phi,\theta,\a_1,\a_2,\a_3)$, which are related to the $y_i$ in the following fashion
\begin{align}
\label{eq:cp3coords}
\nonumber y_1+i y_2 & = r \sin{\theta} \cos{\frac{\a_1}{2}} e^{\frac{i}{2} (\a_2+\a_3) + \frac{i}{2} \phi} = \frac{r}{|w|} w_1, \\
y_3+i y_4 & =  r \sin{\theta} \sin{\frac{\a_1}{2}} e^{-\frac{i}{2} (\a_2-\a_3) + \frac{i}{2} \phi} = \frac{r}{|w|} w_2, \\
\nonumber y_5+i y_6 & = r\cos{\theta} e^{i \phi} = \frac{r}{|w|} w_3,
\end{align}
\noindent where $|w| =\bar{w}_k w_k$ and $\sin{r} = \frac{|w|}{\sqrt{1+|w|^2}}$. Now we would like to effectively extract the angle $\phi$ from this element, (\ref{eq:cpcospar}), in order to write $g_\CP$ as
\[
\nonumber g_\CP = \Lambda_\CP (\phi) \tilde{g}_\CP \, ,
\]
\noindent where $\tilde{g}_\CP$ no longer depends on $\phi$. Inspection of the relations just above (\ref{eq:cp3coords}), shows that extracting $\phi$ corresponds to setting $y_6$ to zero. The remaining five $y$'s in conjunction with $\phi$ still give a good parametrization of $\CP^3$. It is moreover not hard to verify explicitly that on the coset element (\ref{eq:cpcospar}), $T_{34}+T_{56}$ generates shifts in $\phi$, hence giving the form of $\Lambda_\CP (\phi)$.

\smallskip

It is perhaps noteworthy to mention that even though we have introduced $\phi$ in exchange for $y_6$, such that we remain with $y_i$, $i=1,\ldots,5$ and $\phi$, it is important to remember the coordinates are not all completely independent in the usual sense. As an example, confining oneself to the geodesic circle parametrized by $\phi$ corresponds to taking $w_3 = e^{i \phi}$, which means $y_i=0$ for $i=1,\ldots,4$, and $y_5 = \pi/4$.

\smallskip

Lastly the $\Upsilon$ matrix used in the definition of the map $\Omega$ is given by
\begin{eqnarray}
\nonumber \Upsilon=\left(\begin{array}{rl}
K_4 C_4 & 0 \\
0       &  -K_6
\end{array}\right)\, .
\end{eqnarray}
\section{The coset parametrization and $\kappa$-symmetry parameter}\label{kappappendix}

As indicated above, the choice of coset parametrization, does have an effect on the explicit form of the $\kappa$-symmetry parameter. As $\kappa$-symmetry acts on the coset element by multiplication from the right
\[
g\rightarrow g e^\epsilon = g^\prime g_c,
\]
\noindent at linear order in $\chi$ and $\epsilon$, we have the following transformation for our choice of coset parametrization (\ref{eq:cospar})
\[
g\rightarrow g_o g_\chi g_B e^\epsilon = g_o e^\chi e^{g_B \epsilon g_B^{-1}} g_B \approx g_o e^{\chi +g_B \epsilon g_B^{-1}} g_B.
\]
\noindent Thus at the linearized level the fermionic matrix $\chi$ undergoes a shift
\[
\chi \rightarrow \chi +g_B \epsilon g_B^{-1},
\]
\noindent under a $\kappa$-symmetry variation. In the coset parametrization $g=g_\chi g_B$, the light-cone coset element, $g_B \sim \mbox{exp}(\mbox{diag}(i\Gamma^0,T_6)$, commutes with $\epsilon$ as presented in \cite{alday-2008-089}, however for our choice of coset parametrization, we have $g_B  \sim  \mbox{exp}(\mbox{diag}(0,\pi/4 T_5))$ which does not commute with $\epsilon$. As a consequence of this the epsilon parameter needs to be modified by exactly this conjugation by $g_B$. Considering we are interested in gauge choices invariant under the manifest bosonic symmetry, we will also rotate the $\epsilon$ parameter by use of the matrix $S$ (\ref{coordtrans}) which will then clearly indicate some invariant gauge choices. Concretely the parameter of \cite{arutyunov-2008} is calculated from (\ref{eq:kappapar}) by considering a bosonic coset element schematically of the form
\begin{equation*}
A^{(2)}= \left(\begin{array}{cc} i x\Gamma^0 & 0
\\
0 & y T_6
\end{array} \right)\, .
\end{equation*}
\noindent The Virasoro constraint here reads ${\rm
str}(A^{(2)}_{\a,-}A^{(2)}_{\b, -})=0$, and implies $x^2=y^2$. Choosing $x=y$ as a solution and doing the just indicated conjugation and basis change (\ref{coordtrans}), we find
\[
\epsilon^{(1)} \rightarrow S g_B \epsilon^{(1)} (S g_B)^{-1} = \left(
\begin{array}{cc}
0 & \varepsilon \\
 \bullet & 0
\end{array}
\right),
\]
\noindent where
\[
\varepsilon =\frac{1+i}{\sqrt{2}}
\left(
\begin{array}{ccc}
 0 & 0 & \varrho \\
 0 & \sigma_2 \varrho^* \sigma_2  & 0
\end{array}
\right),
\]
\noindent and
\[
\varrho = \left(
\begin{array}{cc}
 i (\theta _{43}+\theta _{44})-\theta _{45}-\theta _{46} & -i (\theta _{43}-\theta _{44})-\theta _{45}+\theta _{46} \\
 -\theta _{43}+\theta _{44}-i (\theta _{45}-\theta _{46}) & -\theta _{43}-\theta _{44}+i (\theta _{45}+\theta _{46})
\end{array}
\right).
\]
\noindent Here the $\theta_{ij}$ correspond to the original entries of the fermionic parameter $\k_{++}$ entering in (\ref{eq:kappapar}). Of course the expression for $\epsilon^{(3)}$ is similar. It should now be clear that this parameter can be used to impose the gauge choice
(\ref{eq:kgaugechoice}).

\smallskip

This expression might appear somewhat asymmetrical, however we note here that picking the alternate solution to the Virasoro constraint in the above, $x=-y$, exactly leads to a parameter schematically of the form
\[
\varepsilon_{\scriptscriptstyle-} =(1+i)\left(
\begin{array}{ccc}
 0 & \varrho_{\scriptscriptstyle-} & 0 \\
 0 & 0 & - \sigma_2 \varrho_{\scriptscriptstyle-}^* \sigma_2
\end{array}
\right).
\]


\begin{thebibliography}{20}
{\small\small

\bibitem{Aharony:2008ug}
  O.~Aharony, O.~Bergman, D.~L.~Jafferis and J.~Maldacena,
  ``N=6 superconformal Chern-Simons-matter theories, M2-branes and their
  gravity duals,''
  JHEP {\bf 0810} (2008) 091
  [arXiv:0806.1218 [hep-th]].

\bibitem{arutyunov-2009}
  G.~Arutyunov and S.~Frolov,
   ``Foundations of the $\AdS$ Superstring. Part I,''
  arXiv:0901.4937 [hep-th].


\bibitem{arutyunov-2008}
  G.~Arutyunov and S.~Frolov,
  ``Superstrings on ${\rm AdS}_4 \times \mathbb{CP}^3$ as a Coset Sigma-model,''
  JHEP {\bf 0809} (2008) 129
  [arXiv:0806.4940 [hep-th]].

\bibitem{Stefanski:2008ik}
  B.~Stefanski~jr.,
  ``Green-Schwarz action for Type IIA strings on $AdS_4\times CP^3$,''
  Nucl.\ Phys.\  B {\bf 808} (2009) 80
  [arXiv:0806.4948 [hep-th]].

\bibitem{Gromov:2008bz}
  N.~Gromov and P.~Vieira,
   ``The ${\rm AdS}_4/{\rm CFT}_3$ algebraic curve,''
  JHEP {\bf 0902} (2009) 040
  [arXiv:0807.0437 [hep-th]].


\bibitem{Gomis:2008jt}
  J.~Gomis, D.~Sorokin and L.~Wulff,
  ``The complete ${\rm AdS}_4 \times \mathbb{CP}^3$ superspace for the type IIA superstring and
  D-branes,''
  JHEP {\bf 0903} (2009) 015
  [arXiv:0811.1566 [hep-th]].

  \bibitem{Grassi:2009yj}
    P.~A.~Grassi, D.~Sorokin and L.~Wulff,
    ``Simplifying superstring and D-brane actions in ${\rm AdS}_4 \times \mathbb{CP}^3$ superbackground,''
    arXiv:0903.5407 [hep-th].

\bibitem{Bonelli:2008us}
  G.~Bonelli, P.~A.~Grassi and H.~Safaai,
  ``Exploring Pure Spinor String Theory on $AdS_4\times \mathbb{CP}^3$,''
  JHEP {\bf 0810} (2008) 085
  [arXiv:0808.1051 [hep-th]].

\bibitem{MZ}
  J.~A.~Minahan and K.~Zarembo,
  ``The Bethe ansatz for superconformal Chern-Simons,''
  JHEP {\bf 0809} (2008) 040
  [arXiv:0806.3951 [hep-th]].

\bibitem{Bak:2008cp}
  D.~Bak and S.~J.~Rey,
  ``Integrable Spin Chain in Superconformal Chern-Simons Theory,''
  JHEP {\bf 0810} (2008) 053
  [arXiv:0807.2063 [hep-th]].

\bibitem{Zwiebel:2009vb}
  B.~I.~Zwiebel,
  ``Two-loop Integrability of Planar N=6 Superconformal Chern-Simons Theory,''
  arXiv:0901.0411 [hep-th].

\bibitem{Minahan:2009te}
  J.~A.~Minahan, W.~Schulgin and K.~Zarembo,
  ``Two loop integrability for Chern-Simons theories with N=6 supersymmetry,''
  JHEP {\bf 0903} (2009) 057
  [arXiv:0901.1142 [hep-th]].

\bibitem{Bak:2009mq}
  D.~Bak, H.~Min and S.~J.~Rey,
  ``Generalized Dynamical Spin Chain and 4-Loop Integrability in N=6
  Superconformal Chern-Simons Theory,''
  arXiv:0904.4677 [hep-th].

\bibitem{Gromov:2008qe}
  N.~Gromov and P.~Vieira,
  ``The all loop AdS4/CFT3 Bethe ansatz,''
  JHEP {\bf 0901} (2009) 016
  [arXiv:0807.0777 [hep-th]].


\bibitem{McLoughlin:2008ms}
  T.~McLoughlin and R.~Roiban,
  ``Spinning strings at one-loop in ${\rm AdS}_4 \times \mathbb{P}^3$,''
  JHEP {\bf 0812} (2008) 101
  [arXiv:0807.3965 [hep-th]].


\bibitem{alday-2008-089}
  L.~F.~Alday, G.~Arutyunov and D.~Bykov,
   ``Semiclassical Quantization of Spinning Strings in ${\rm AdS}_4 \times \mathbb{CP}^3$,''
  JHEP {\bf 0811} (2008) 089
  [arXiv:0807.4400 [hep-th]].

\bibitem{Krishnan:2008zs}
  C.~Krishnan,
  ``AdS4/CFT3 at One Loop,''
  JHEP {\bf 0809} (2008) 092
  [arXiv:0807.4561 [hep-th]].

  \bibitem{sundin-2008}
    P.~Sundin,
    ``The ${\rm AdS}_4 \times \mathbb{CP}^3$ string and its Bethe equations in the near plane wave
    limit,''
    JHEP {\bf 0902} (2009) 046
    [arXiv:0811.2775 [hep-th]].

  \bibitem{uvarov-2008}
    D.~V.~Uvarov,
     ``${\rm AdS}_4 \times \mathbb{CP}^3$ superstring and $D=3$ $N=6$ superconformal symmetry,''
    arXiv:0811.2813 [hep-th].


\bibitem{Bergman:2009zh}
  O.~Bergman and S.~Hirano,
  ``Anomalous radius shift in ${\rm AdS}_4/{\rm CFT}_3$,''
  arXiv:0902.1743 [hep-th].


  \bibitem{Bykov:2009jy}
    D.~Bykov,
    ``Off-shell symmetry algebra of the ${\rm AdS}_4 \times \mathbb{CP}^3$ superstring,''
    arXiv:0904.0208 [hep-th].

  \bibitem{Zarembo:2009au}
    K.~Zarembo,
    ``Worldsheet spectrum in ${ \rm AdS}_4/{\rm CFT}_3$ correspondence,''
    arXiv:0903.1747 [hep-th].

\bibitem{Kalousios:2009ey}
  C.~Kalousios, C.~Vergu and A.~Volovich,
  ``Factorized Tree-level Scattering in ${ \rm AdS}_4/{\rm CFT}_3$,''
  arXiv:0905.4702 [hep-th].
  
\bibitem{Grignani:2008is}
  G.~Grignani, T.~Harmark and M.~Orselli,
  ``The SU(2) $\times$ SU(2) sector in the string dual of N=6 superconformal
  Nucl.\ Phys.\  B {\bf 810} (2009) 115
  [arXiv:0806.4959 [hep-th]].

\bibitem{Astolfi:2008ji}
  D.~Astolfi, V.~G.~M.~Puletti, G.~Grignani, T.~Harmark and M.~Orselli,
  ``Finite-size corrections in the SU(2) $\times$ SU(2) sector of type IIA string
  theory on ${\rm AdS}_4 \times \mathbb{CP}^3$,''
  Nucl.\ Phys.\  B {\bf 810} (2009) 150
  [arXiv:0807.1527 [hep-th]].

  \bibitem{Rashkov:2008rm}
    R.~C.~Rashkov,
    ``A note on the reduction of the ${\rm AdS}_4 \times \mathbb{CP}^3$ string sigma model,''
    Phys.\ Rev.\  D {\bf 78} (2008) 106012
    [arXiv:0808.3057 [hep-th]].


\bibitem{AF}
  G.~Arutyunov and S.~Frolov,
  ``Integrable Hamiltonian for classical strings on $\AdS$,''
  JHEP {\bf 0502} (2005) 059
  [arXiv:hep-th/0411089].


  \bibitem{Alday:2005jm}
    L.~F.~Alday, G.~Arutyunov and S.~Frolov,
     ``New integrable system of 2dim fermions from strings on $\AdS$,''
    JHEP {\bf 0601} (2006) 078
    [arXiv:hep-th/0508140].



\bibitem{Arutyunov:2005hd}
  G.~Arutyunov and S.~Frolov,
  ``Uniform light-cone gauge for strings in $\AdS$: Solving $\alg{su}(1|1)$
  JHEP {\bf 0601} (2006) 055
  [arXiv:hep-th/0510208].


}
\end{thebibliography}
\end{document}